\documentclass{rmf-d}
\usepackage{nopageno,rmfbib,multicol,times,epsf,amsmath,amssymb,cite}
\usepackage[latin1]{inputenc}
\usepackage[]{caption2}
\usepackage{graphicx}
\usepackage{siunitx}
\usepackage{hyperref}
\usepackage{lineno}
\def\rmfcornisa{Research OR Education of Physics \hfill\rmf\ {\bf ??} (*?*) ???--???
\hfill MES? A\~NO?}

\def\rmfcintilla{{\it Rev.\ Mex.\ Fis.\/} {\bf ??} (*?*) (????) ???--???}
\clearpage \rmfcaptionstyle 
\begin{document}
\title{   Accessing glue through photoproduction measurements at GlueX 
\vspace{-6pt}}
\author{ Peter Pauli  }
\address{ University of Glasgow, Glasgow G12 8QQ, United Kingdom  }
\author{ for the GlueX collaboration }
\maketitle
\recibido{day month year}{day month year
\vspace{-12pt}}
\begin{abstract}
\vspace{1em} 
Photoproduction experiments are a key tool in the investigation of the spectrum of hadronic states and the way gluons contribute to this spectrum. The GlueX experiment, located at Jefferson Lab, features a linearly polarized tagged photon beam and its detector system is optimized to measure a wide range of neutral and charged final states. GlueX offers unique capabilities to study the spectrum of hadrons and is dedicated to the search for hybrid mesons, states with gluonic degrees of freedom.
This talk presents first results from our initial campaign of data taking which finished in 2018.
\vspace{1em}
\end{abstract}
\keys{  GlueX; SDME; hybrid mesons; pentaquark  \vspace{-4pt}}
\pacs{   \bf{\textit{25.20.-x, 25.20.Lj}}    \vspace{-4pt}}
\begin{multicols}{2}

\section{Introduction}\label{sec:intro}
The field of hadron spectroscopy provides vital input for our understanding of Quantum Chromodynamics (QCD), the theory describing the strong force. Photoproduction measurements play a key role in this as they provide unique insights into the production of many meson and baryon states. The GlueX experiment aims to measure exotic hybrid mesons. These mesons do not consist solely of a $q\bar{q}$ pair but also contain a gluon which interacts in such a way that it contributes to the overall quantum numbers of the state. Therefore, hybrid mesons are no longer constrained to quantum numbers of
\[ \vec{J} = \vec{L}+\vec{S}, \,\, P=(-1)^{L+1}, \,\,C = (-1)^{L+S}, \] where $\vec{J}$, $\vec{L}$, $\vec{S}$ are the overall spin, angular momentum and intrinsic spin, respectively, and $P$ and $C$ are parity and charge conjugation parity of the meson. They can also have quantum numbers such as $1^{-+}$. These quantum numbers, which can not be obtained in a simple quark model picture, are considered a {\it smoking gun} in the search for hybrid mesons. Over the course of the last couple of years, various experiments claimed evidence for two such $1^{-+}$ states, commonly called $\pi_1(1400)$ and $\pi_1(1600)$ \cite{IHEP-Brussels-LosAlamos-AnnecyLAPP:1988iqi,Aoyagi:1993kn,VES:1993scg,E852:1997gvf,CrystalBarrel:1998cfz,CrystalBarrel:1999reg,COMPASS:2014vkj,VES:1992zkx,Dorofeev:1999th,E852:1998mbq,Chung:2002pu,E852:2001ikk,COMPASS:2009xrl}. Recently, the JPAC collaboration published a coupled-channel analysis of COMPASS data in which they show that they need only one pole, i.e. one $\pi_1$ state, to describe the data for $\pi_1(1400)\rightarrow\eta\pi$ and $\pi_1(1600)\rightarrow\eta'\pi$ with a mass of about \SI{1.56}{\GeV} and a width of about \SI{0.49}{\GeV} \cite{JPAC:2018zyd} . Consequently, GlueX is initially focussing on the $\eta^{(\prime)}\pi$ channels, which can each be measured via multiple decay modes.\par
The GlueX experimental setup (see Figure \ref{fig:detector}) consists of a detector with almost full acceptance, designed to measure a wide range of charged and neutral final states. A detailed description is given in Reference \cite{GlueX:2020idb} and briefly summarized here. The electron beam delivered by Jefferson Lab's CEBAF electron accelerator is converted into a linearly polarized photon beam using the coherent bremsstrahlung technique on a thin diamond radiator \cite{Timm:1969mf}. In its coherent peak, which is located at around \SI{9}{\GeV} photon energy, the degree of polarization reaches $\sim40\%$. The photon beam, which travels for \SI{75}{m} before being collimated, is incident on a LH$_2$ target. The target is surrounded by the Start Counter, Central Drift Chamber and Barrel Calorimeter. These, together with a second drift chamber in the forward direction, are enclosed in a \SI{2}{\tesla} solenoidal magnetic field. Outside the magnetic field in the forward direction are a Time-of-Flight wall and a second calorimeter. This combination of different detector types achieves good momentum and energy resolutions for charged and neutral particles and is key to GlueX's ability to measure a wide variety of final states. \par
GlueX's first campaign of data-taking ran from 2017-2018 and the following results are based on this data set or subsets of it. 
\begin{center}
	\includegraphics[width=\columnwidth]{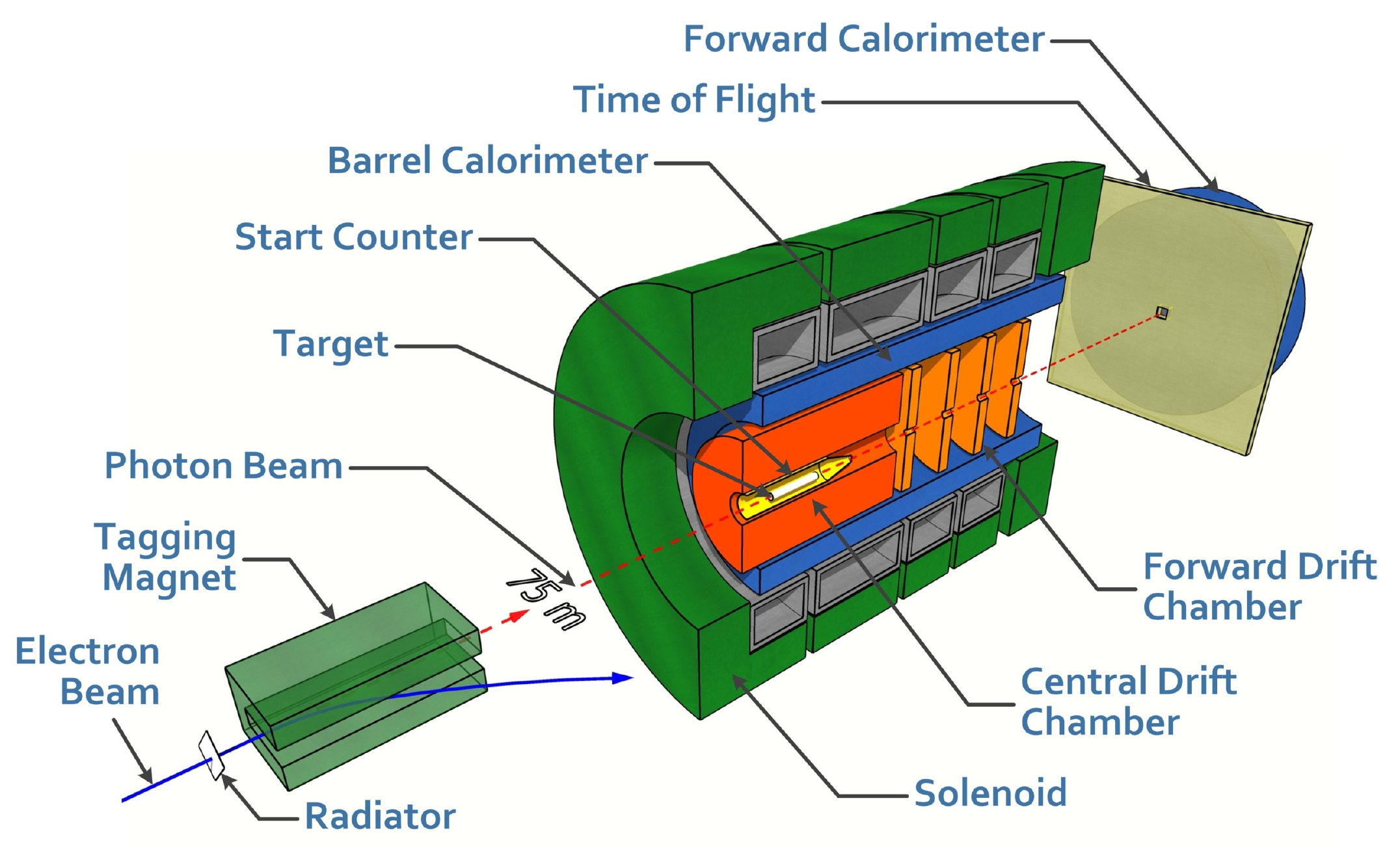} \\
	\refstepcounter{figure}
	\rmfcaptionstyle{Figure \ref{fig:detector}: Overview over the GlueX detector (taken from Reference \cite{GlueX:2020idb}).}\label{fig:detector} \\
\end{center}

\section{Spin-density matrix elements}\label{sec:sdme}
In a first step towards a full partial-wave analysis (PWA), which is necessary to establish the existence of a state like the $\pi_1$, GlueX measures polarization observables such as spin-density matrix elements (SDMEs) for a wide range of states. These SDMEs contain information about the spin polarization of the produced state and so they can be used to learn about the production processes involved in photoproduction at GlueX energies, an energy regime which is mostly unexplored. In order to obtain SDMEs the angular distributions of the decay particles of a state are measured. These can then be fitted with an intensity function from which all accessible SDMEs can be extracted. Parameter estimation is performed using the unbinned maximum likelihood technique, similar to what is done in a PWA. As such, we can exercise our analysis toolkits and make sure that everything performs as expected before moving on to full PWAs. In the case of the reaction $\gamma p\rightarrow \Lambda(1520)K^+ \rightarrow pK^-K^+$ the intensity function is given by Equation \eqref{eq:SDMEfunction} \cite{Yu:2017kng}:
\end{multicols}
\medline
    \begin{align}
    	W(\theta,\phi,\Phi)	&= \frac{1}{2\pi} \frac{d\sigma}{dt}\frac{3}{4\pi}\left\{\rho^0_{33}\sin^2{\theta}+ \rho^0_{11}\left(\frac{1}{3}+\cos^2{\theta}\right) - \frac{2}{\sqrt{3}}\text{Re}\rho^0_{31}\sin{2\theta}\cos{\phi} - \frac{2}{\sqrt{3}}\text{Re}\rho^0_{3-1}\sin^2{\theta}\cos{2\phi}\right.\nonumber\\
    	&- P_\gamma\cos{2\Phi} \left[\rho^1_{33}\sin^2{\theta}+ \rho^1_{11}\left(\frac{1}{3}+\cos^2{\theta}\right) - \frac{2}{\sqrt{3}}\text{Re}\rho^1_{31}\sin{2\theta}\cos{\phi} - \frac{2}{\sqrt{3}}\text{Re}\rho^1_{3-1}\sin^2{\theta}\cos{2\phi}\right]\nonumber\\
    	&- P_\gamma\sin{2\Phi} \frac{2}{\sqrt{3}}\left[\text{Im}\rho^2_{31}\sin{2\theta}\sin{\phi} + \text{Im}\rho^2_{3-1}\sin^2{\theta}\sin{2\phi}\right]\left.\vphantom{\frac{1}{2}}\right\}. \label{eq:SDMEfunction}
    \end{align}
\medline
\begin{multicols}{2}
\noindent The angles $\theta$ and $\phi$ denote the angles of the $K^-$ in the Gottfried-Jackson frame \cite{GJ:1964}. $\Phi$ is the angle between the photon polarization plane and the hadronic production plane and $P_\gamma$ is the degree of linear polarization of the photon. The SDMEs are denoted as $\rho^i_{2\lambda_\Lambda,2\lambda'_\Lambda}$, with $\lambda_\Lambda$ denoting the $\Lambda(1520)$ helicity, and $i=0$ denoting unpolarized SDMEs while $i=1,2$ denotes polarized SDMEs. The differential cross-section $\frac{d\sigma}{dt}$ assures proper normalization such that $\rho^0_{11}+\rho^0_{33} = \frac{1}{2}$ \cite{GlueX:2021pcl}. Assuming a t-channel production process in which a particle $X$ is exchanged between the photon and target proton, one can study the {\it naturality} $\eta=P(-1)^J$ of $X$, where $P$ denotes its parity and $J$ its spin. A positive naturality (natural exchange) implies that $X$ is e.g. a vector or tensor meson, while negative naturality (unnatural exchange) implies that $X$ is e.g. a pseudo-scalar or axial-vector meson \cite{GlueX:2021pcl}. It is possible to express purely natural ($N$) and unnatural ($U$) production amplitudes as linear combinations of SDMEs, e.g. $\rho^{0}_{11} + \rho^{1}_{11} = \frac{2}{\mathcal{N}}\left(|N_{0}|^{2}+|N_{1}|^{2}\right)$, with normalisation $\mathcal{N}$ (see Reference \cite{GlueX:2021pcl} for a derivation and detailed description of these combinations). Figure \ref{fig:lambda} shows results obtained by GlueX for different combinations of purely natural or purely unnatural production amplitudes binned in four-momentum transfer from the photon to the target proton. Across the whole momentum transfer region, natural amplitudes dominate over unnatural ones. In fact, unnatural amplitudes are mostly consistent with 0 and only show some small contributions for vanishing four-momentum transfer. This behaviour is predicted by a model calculation performed by Yu and Kong \cite{Yu:2017kng} which is shown as solid blue and red dotted lines in Figure \ref{fig:lambda}. Their calculation shows a dominance of natural amplitudes but the relative strength of the different combinations of SDMEs is not reproduced by the data. This is not unexpected since these are the first measurements of unpolarized SDMEs for this reaction at these energies and the first measurements of polarized SDMEs for this reaction at all.
\begin{center}
	\includegraphics[width=0.9\columnwidth]{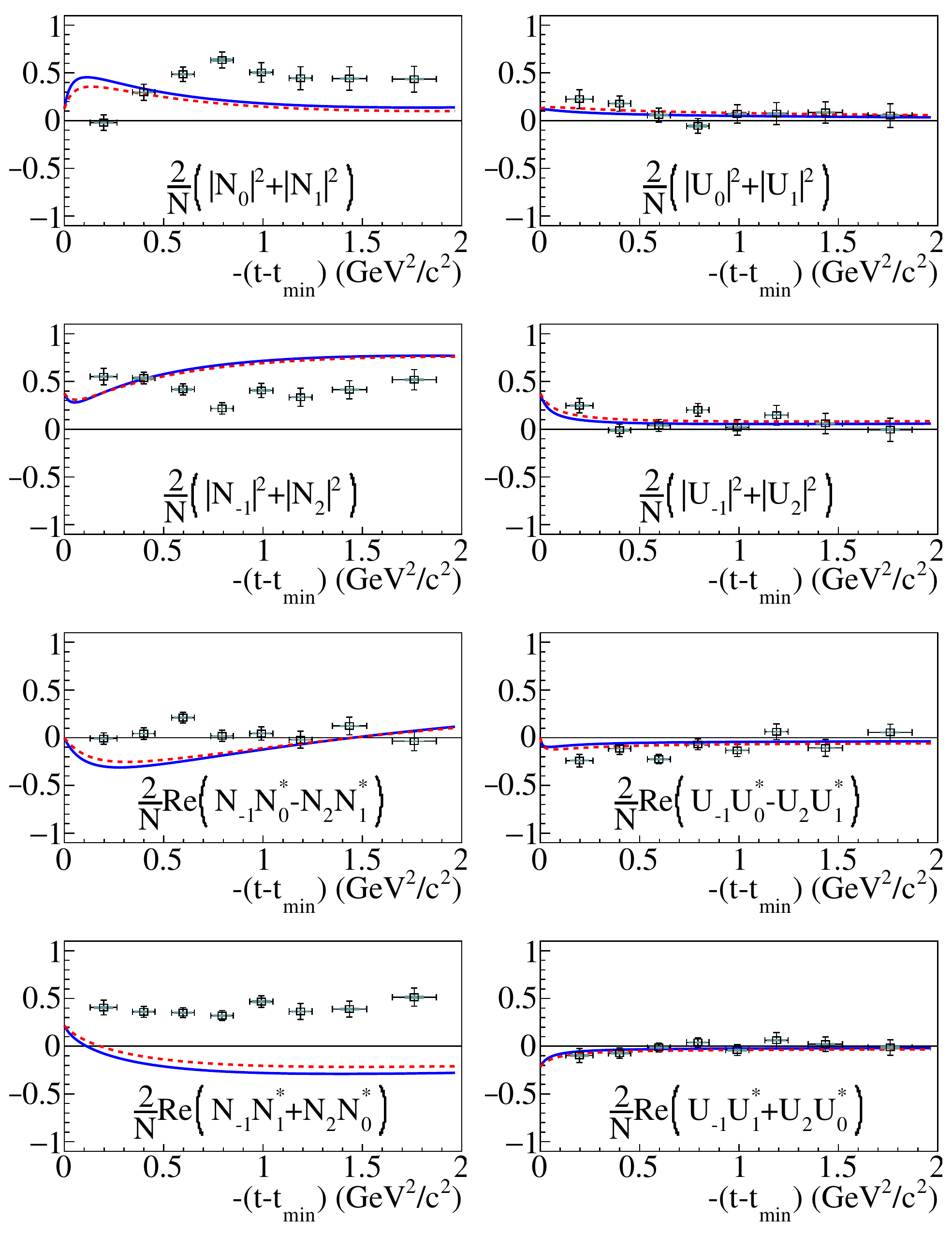} \\
	\refstepcounter{figure}
	\rmfcaptionstyle{Figure \ref{fig:lambda}: Combinations of SDMEs representing purely natural ($N$) and purely unnatural ($U$) production amplitudes for the reaction $\gamma p\rightarrow \Lambda(1520)K^+ \rightarrow pK^-K^+$ (taken from \cite{GlueX:2021pcl}). The solid blue and red dotted lines represent model calculations performed by Yu and Kong \cite{Yu:2017kng}.} \label{fig:lambda}
	\vspace{0.cm}
\end{center}\par
Similar analyses are performed for other reaction channels such as $\rho(770)$, $\omega(782)$ or $\phi(1020)$ photoproduction. These types of analyses not only provide valuable physics information, used to make informed decisions regarding wave sets for PWA, but also help us to identify problems in the GlueX detector simulation.  PWA and SDME measurements both rely on acceptance corrections, obtained from Monte Carlo simulations using a Geant4 \cite{Agostinelli2003} based description of our detector system. Careful SDME analyses in a variety of final states help us to improve the agreement between data and simulation.

\section{Hybrid search in $\eta\pi$}\label{sec:hybrid}
While production mechanisms are being established for a range of different final states we are starting to perform partial-wave analyses in the $\eta\pi$ channel. A more detailed account of the ongoing efforts is given elsewhere in these proceedings \cite{CG:hadron} but in order to illustrate the analysis approach and formalism we will focus on the reactions $\gamma p\rightarrow\eta\pi^0 p\rightarrow4\gamma p$ and $\gamma p\rightarrow\eta\pi^- \Delta^{++}\rightarrow2\gamma \pi^-\pi^+p$. In these final states one expects to find the two well established $a_0(980)$ and $a_2(1320)$ meson resonances, with $J^{PC}$ quantum numbers $0^{++}$ and $2^{++}$, respectively. The $a_0(980)$ is expected to decay in S-wave and the  $a_2(1320)$ in D-wave into $\eta\pi$, which are both pseudoscalars. Confirming these expectations in an initial PWA will show that the formalism works and that experimental effects such as detector acceptance are well under control.\par
The formalism used to analyze the data uses polarized photoproduction amplitudes \cite{PhysRevD.100.054017}. The intensity function used to fit the data is given by Equation \eqref{eq:PWAfunction}:
\end{multicols}
\vspace{-0.5cm}\medline
    \begin{align}
    	I(\theta,\phi,\Phi)= 2\kappa\sum_k\left\{ \vphantom{\frac{1}{2}} \right. &  \left(1-P_\gamma\right)\left|\sum_{lm}[l]^{(-)}_{m;k}\text{Re}[Z_l^m(\theta,\phi,\Phi)]\right|^2 + \left(1-P_\gamma\right)\left|\sum_{lm}[l]^{(+)}_{m;k}\text{Im}[Z_l^m(\theta,\phi,\Phi)]\right|^2 \nonumber\\
													& \left.  \left(1+P_\gamma\right)\left|\sum_{lm}[l]^{(+)}_{m;k}\text{Re}[Z_l^m(\theta,\phi,\Phi)]\right|^2 + \left(1+P_\gamma\right)\left|\sum_{lm}[l]^{(-)}_{m;k}\text{Im}[Z_l^m(\theta,\phi,\Phi)]\right|^2 \vphantom{\frac{1}{2}}\right\} \label{eq:PWAfunction}
    \end{align}
\medline
\begin{multicols}{2}
\noindent The kinematic factors are summarised in $\kappa$, $P_\gamma$ and $\Phi$ are defined as in Equation \eqref{eq:SDMEfunction} and $\theta$ and $\phi$ are the angles of the $\eta$ in the helicity frame. Also we have $Z_l^m(\theta,\phi,\Phi)=Y_l^m(\theta,\phi)e^{-i\Phi}$ with $Y_l^m(\theta,\phi)$ being spherical harmonics. The spin quantum number and its projection for each wave are given by $l$ and $m$. The fit parameters extracted from the fit are $[l]^{(\epsilon)}_{m;k}$, where $\epsilon$ denotes the reflectivity of the wave which corresponds to the previously defined naturality in GlueX's energy range, and $k$ denotes the nucleon helicity, which is not explicitly measured in GlueX. The initial wave set used in the preliminary fits presented here is $S_0^\pm$, $D_{-1,0,1}^\pm$, $D_2^+$. This wave set is based on a tensor meson photoproduction model by JPAC \cite{PhysRevD.102.014003}.\par
The fit results for the dominant waves are presented in Figure \ref{fig:etapi} in bins of $M(\eta\pi)$. The data covers the four-momentum transfer range from $-t=\SIrange[range-phrase=-]{0.1}{0.3}{\GeV^2}$. The black data points show the total amount of data available in this reaction. The red and blue data points show the contributions from $D^+_1$ and $D^-_1$, respectively, the green data points show the $D^+_2$ contributions. The $a_2(1320)$ is expected in a D-wave. For $\eta\pi^0$ we can clearly see it appear in the $D_2^+$ wave (Figure \ref{fig:etapi}, top) . This dominance of a positive reflectivity wave corresponds to a dominance of natural production amplitudes, such as $\rho$ or $\omega$ exchange. Production of the $a_2(1320)$ in a $m=2$ helicity state in this channel is consistent with what was seen by Belle in $\gamma\gamma\rightarrow\eta\pi$ \cite{PhysRevD.80.032001}. For $\eta\pi^-$ on the other hand the $a_2(1320)$ appears dominantly in a $D_1^-$ wave in this range of $-t$ (Figure \ref{fig:etapi}, bottom). This corresponds to unnatural production amplitudes such as $\pi$ exchange. There is also a small $D_1^+$ contribution visible which indicates that the production proceeds through more than just one exchange mechanism. In both channels we observe a small signal around the $a_2(1700)$ mass. The signal and its phase motions are currently under study as they are critical for the hybrid meson search in these channels. \par
These preliminary results are very encouraging and show that GlueX is well on track to perform a full partial-wave analysis for $\eta^{(\prime)}\pi$ with high statistical precision. Compared to results from previous experiments, GlueX has the advantage of being able to measure $\eta^{(\prime)}\pi$ in multiple final states. This will help us improve our statistics as well as aid in uncovering and controlling systematic uncertainties. Compared to previous results by COMPASS \cite{COMPASS:2014vkj}, GlueX collected similar amounts of data for $\eta\pi^-$ with $\eta\rightarrow\pi^+\pi^-\pi^0$ alone and about a factor ten more in $\eta\rightarrow\gamma\gamma$. The full partial-wave analysis will contain the exotic $1^{-+}$ waves which were seen by various experiments, as introduced earlier. Results from this analysis are greatly anticipated and will provide crucial information for our understanding of hybrid mesons. This constitutes GlueX's first step towards establishing a spectrum of exotic hybrid mesons and other analyses are ongoing to further the search.\par
\begin{center}
	\includegraphics[width=0.9\columnwidth,trim={5cm 4.5cm 6cm 5cm}, clip]{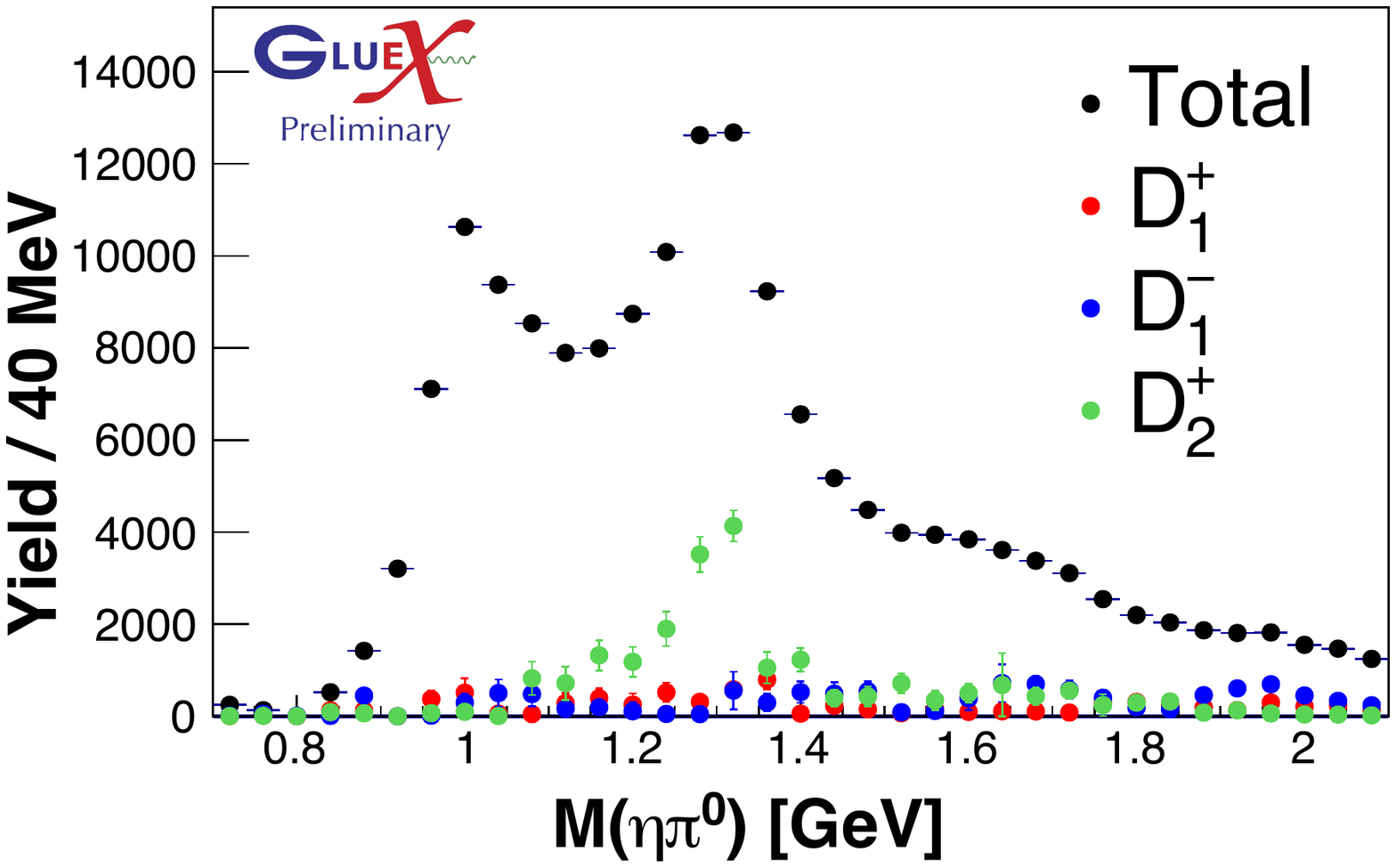}\\
	\includegraphics[width=0.9\columnwidth,trim={5cm 4.5cm 6cm 5cm}, clip]{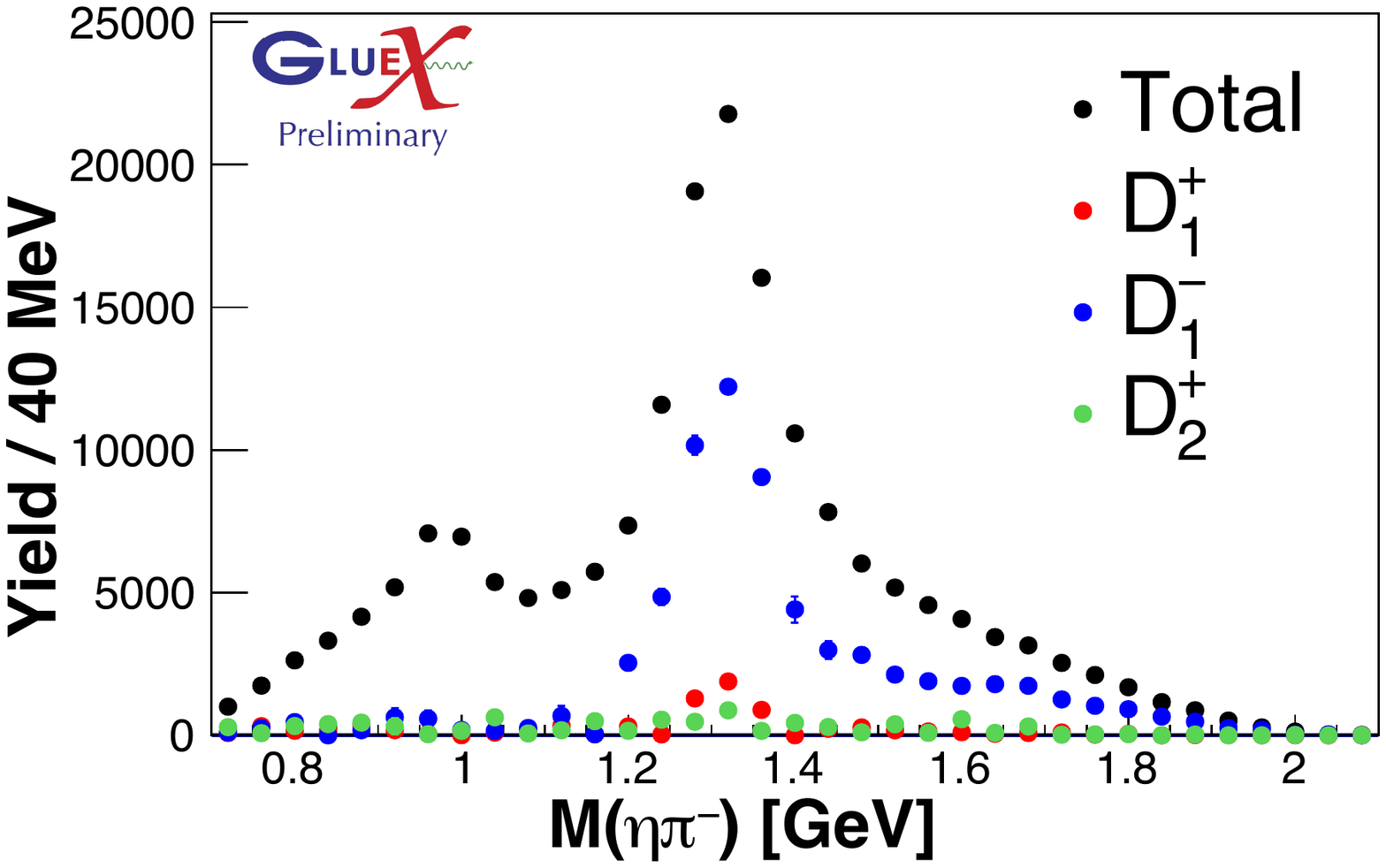}\\
	\refstepcounter{figure}
	\rmfcaptionstyle{Figure \ref{fig:etapi}: Preliminary results of the partial-wave analysis of $\gamma p\rightarrow\eta\pi^0 p\rightarrow4\gamma p$ (top) and $\gamma p\rightarrow\eta\pi^- \Delta^{++}\rightarrow2\gamma \pi^-\pi^+p$ (bottom). The chosen wave set is based on a tensor meson photoproduction model \cite{PhysRevD.102.014003}. Shown are only the dominant D-wave contributions.}\label{fig:etapi}\\
\end{center}

\section{$J/\psi$ cross-section near threshold}\label{sec:jpsi}
The photoproduction of $J/\psi$ mesons close to threshold of $E_\gamma=\SI{8.2}{\GeV}$ is expected to hold valuable information on the gluonic content of the proton. So far only two experiments published data in this region and both were performed in the 1970s \cite{Gittelman:1975ix,Camerini:1975cy}. This in itself makes the measurement of this process at GlueX very desirable. Additional motivation was provided by LHCb's announcement in 2015, when they provided evidence for two potential pentaquarks decaying into $J/\psi p$ \cite{LHCb:2015yax}, a very broad $P_c^+(4380)$ and a narrower $P_c^+(4450)$, with preferred spins of $\frac{3}{2}$ and $\frac{5}{2}$ and opposite parity. In 2019 they published an updated analysis which shows that there are potentially three pentaquarks near the $J/\psi p$ threshold \cite{LHCb:2019kea}. The $P_c^+(4380)$ was not found anymore but a $P_c^+(4312)$ was found and the $P_c^+(4450)$ was resolved into two peaks, $P_c^+(4440)$ and $P_c^+(4457)$. The new analysis was not based on a partial-wave analysis and no spin assignments were suggested. These results drew a lot of attention but the nature of the observed structures has not been unambiguously determined. There are several explanations ranging from compact pentaquarks ({\it e.g.} \cite{Maiani:2015vwa,Zhu:2015bba}), to molecular states ({\it e.g.} \cite{Chen:2015loa,Roca:2015dva}) or rescattering effects producing resonance-like structures ({\it e.g.} \cite{Mikhasenko:2015vca,Guo:2015umn}). GlueX can produce these potential pentaquark states directly through s-channel production, a process free of rescattering effects. Pentaquark states could then be observed as structures in the cross-section.\par
In GlueX we can study $J/\psi$ production in an exclusive measurement where $\gamma p\rightarrow J/\psi p\rightarrow e^+e^-p$. The $J/\psi$ can then be identified as a narrow peak in the $e^+e^-$ invariant mass at $M(e^+e^-)=\SI{3.096\pm0.001}{\GeV}$. Figure \ref{fig:jpsi_im} shows the invariant mass spectrum, with an inset showing the $J/\psi$ region at a larger scale. The background under the $J/\psi$ peak is very small and the data were fitted with a linear polynomial and a Gaussian distribution. In total $469\pm22$ $J/\psi$'s were identified.\par
\begin{center}
	\includegraphics[width=0.9\columnwidth]{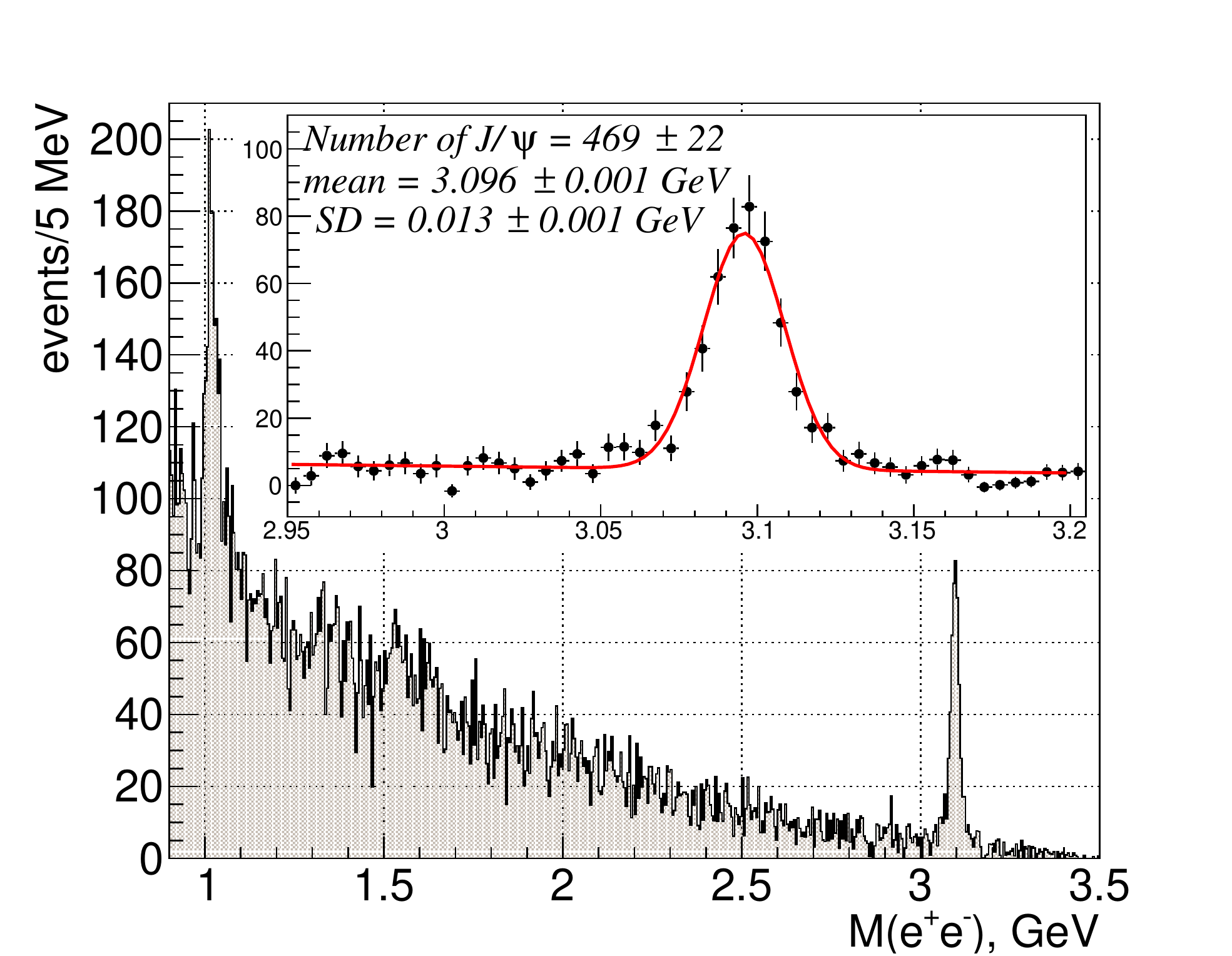} \\
	\refstepcounter{figure}
	\rmfcaptionstyle{Figure \ref{fig:jpsi_im}: Invariant mass distribution for the $e^+e^-$ system. At low masses the peak from the decay of $\phi\rightarrow e^+e^-$ is visible. The $J/\psi$ is visible as a prominent peak at \SI{3.096\pm0.001}{\GeV}. The inset figure shows the axis expanded about the $J/\psi$ mass and includes a fit with a linear polynomial and a Gaussian function. Taken from \cite{GlueX:2019mkq}.}\label{fig:jpsi_im}\\
\end{center}\par
In order to determine the cross-section the results were normalized to the Bethe-Heitler (BH) cross-section between \SIrange[range-phrase=-]{1.2}{2.5}{\GeV}, the continuum between the $\phi$ and $J/\psi$ peaks, which is dominated by the BH process. This normalization avoids systematic uncertainties from luminosity measurements and cancels common acceptance effects. The remaining systematic uncertainty on the absolute scaling is $27\%$. The results for the total cross-section are shown in Figure \ref{fig:jpsi}. The GlueX results are shown as black circles and compared to data points from SLAC in red squares \cite{Camerini:1975cy} and Cornell in blue triangles \cite{Gittelman:1975ix}. Also shown in the plot are models by Kharzeev et all (black dashed) \cite{Kharzeev:1999jt} and Brodsky et al (red dashed) \cite{Brodsky:2000zc} attempting to describe $J/\psi$ photoproduction near threshold. The two contributions in the Brodsky model are scaled such that the incoherent sum describes the data well. According to this model the $J/\psi$ production at threshold is best described by three-gluon-exchange, i.e. all three constituent quarks are involved in the $J/\psi$ production, while at higher energies two-gluon-exchange takes over, i.e. only two constituent quarks are involved. The blue line shows a JPAC model used to set a model-dependent upper limit on the production of the $P_c^+(4440)$ pentaquark candidate \cite{HillerBlin:2016odx}. The underlying assumption is that the $J/\psi$ photoproduction can be described through vector meson dominance. If this is the case, the production and decay of a $P_c^+$ pentaquark is governed by the branching ratio BR$(P_c^+\rightarrow J/\psi p)$. Making this assumption and treating all $P_c^+$ states as $\frac{3}{2}^-$ states, the cross-section shown in Figure \ref{fig:jpsi} can be used to set upper limits of BR$(P_c^+(4312)\rightarrow J/\psi p)=4.6\%$,  BR$(P_c^+(4440)\rightarrow J/\psi p)=2.3\%$ and BR$(P_c^+(4457)\rightarrow J/\psi p)=3.8\%$. These upper limits include not only the uncertainties on individual data points but also take into account the overall normalization uncertainty as well as uncertainties from the non-resonant parametrization and Breit-Wigner parameter \cite{GlueX:2019mkq} (suppl. material).\par
The data published in Reference \cite{GlueX:2019mkq} and presented here represent about $25\%$ of the GlueX-I data. The remaining data are currently under analysis and results are expected soon. The improved statistical precision will make it possible to report differential cross-sections in bins of momentum transfer $t$ and photon energy $E_\gamma$. These measurements will shed more light on the existence of pentaquarks as well as provide more valuable data on the gluonic content of the proton.
\begin{center}
	\includegraphics[width=0.95\columnwidth]{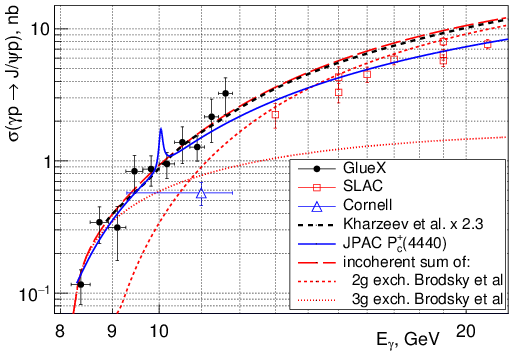} \\
	\refstepcounter{figure}
	\rmfcaptionstyle{Figure \ref{fig:jpsi}: Total cross-section results for $\gamma p\rightarrow J/\psi p$. The GlueX results (black circles) are compared to results from SLAC (red squares) \cite{Camerini:1975cy} and Cornell (blue triangle) \cite{Gittelman:1975ix}. Also shown are models by Kharzeev et all (black dashed) \cite{Kharzeev:1999jt} and Brodsky et al (red dashed) \cite{Brodsky:2000zc} attempting to describe $J/\psi$ photoproduction near threshold. The blue line shows a JPAC model with BR$(P_c^+(4440)\rightarrow J/\psi p)=1.6\%$, used to set an upper limit on the production of the $P_c^+(4440)$ pentaquark candidate \cite{HillerBlin:2016odx}. Taken from \cite{GlueX:2019mkq}.}\label{fig:jpsi}\\
\end{center}

\section{Summary and Outlook}\label{sec:summary}
In this talk we have presented the latest results from GlueX, showing that we are making good progress towards our main goal of studying exotic hybrid mesons. We presented the necessary steps taken on the way to a successful partial-wave analysis that can identify exotic waves in the $\eta^{(\prime)}\pi$ channel. We measure polarization observables such as spin-density matrix elements for a variety of different reactions to determine the dominant production processes at GlueX photon energies. In the case of $\Lambda(1520)$ production, the SDMEs point towards a dominance of natural amplitudes - a result we also see in other reactions. Our preliminary partial-wave analysis in the $\eta\pi$ channel also shows this dominance of natural amplitudes. We can identify S- and D-waves in positive reflectivity corresponding to the known states $a_0(980)$ and $a_2(1320)$. \par
Beyond our ongoing search for exotic hybrid mesons, we presented results for $J/\psi$ photoproduction near threshold. These results inform our understanding of the gluonic content of the proton but also provide us with a way to search for the pentaquark candidates reported by LHCb. Using about $25\%$ of the data, no signal was observed in the total cross-section and model-dependent upper limits were set. The remaining data are currently under analysis and results are expected soon. \par
Since 2019, GlueX has been taking data with an additional detector. The DIRC (detection of internally reflected Cherenkov light), which is installed in the forward direction between the solenoid and the TOF wall, will greatly enhance GlueX's pion-kaon separation. This will enable us to extend our search for exotic mesons to the strangeness sector.\par
GlueX is underway to deliver on its scientific goals and many analyses are currently being carried out. Exciting results can be expected in the near future.

\section{Acknowledgments}
This material is based upon work supported by the U.S. Department of Energy, Office of Science, Office of Nuclear Physics under contract DE-AC05-06OR23177 \footnote{\url{gluex.org/thanks}}. This work was supported by the UK Science and Technology Facilities Council.

\end{multicols}
\medline
\begin{multicols}{2}

\end{multicols}
\end{document}